\documentclass[intlimits,twoside,a4paper]{article}

\usepackage{amsmath,amssymb}
\usepackage{graphicx}
\usepackage{dcolumn}
\usepackage{color}
\usepackage{subfigure}

\usepackage[T2A]{fontenc}
\usepackage[cp1251]{inputenc}

\newcommand{\reffig}[1]{figure~\ref{#1}}

\newcommand{\refsec}[1]{section~\ref{#1}}
\newcommand{\refcite}[1]{reference~\cite{#1}}
\newcommand{\payoff}{\text{payoff}} 

\usepackage[eqsecnum]{cmpj2}

\issue{2014}{17}{3}{33001}
\doinumber{10.5488/CMP.17.33001}

\title[IPD limited attention]
{Iterated Prisoners Dilemma with limited attention
}
\author[U. \c{C}etin, H.O. Bingol]{
	U. \c{C}etin\refaddr{label1,label2},
	H.O. Bingol\refaddr{label1}}
\addresses{
	\addr{label1} Department of Computer Engineering,
	Bogazici University, 34342 Bebek, Istanbul, Turkey
	\addr{label2} Department of Computer Engineering,
	Istanbul Gelisim University, 34310 Avcilar, Istanbul, Turkey
}

\date{Received April 30, 2014}
\authorcopyright{U. \c{C}etin, H.O. Bingol, 2014}

\begin{document}

\maketitle

\begin{abstract}
How attention scarcity effects the outcomes of a game?
We present our findings
on a version of the
Iterated Prisoners Dilemma (IPD) game
in which players can
accept or refuse to play with their partner.
We study the memory size effect on determining the right partner to interact with.
We investigate the conditions under which
the cooperators are more likely to be advantageous than the defectors.
This work demonstrates that,
in order to beat defection,
players do not need a full memorization of each action of all opponents.
There exists a critical attention capacity threshold to beat defectors.
This threshold depends not only on
the ratio of the defectors in the population
but also on the attention allocation strategy of the players.
\pacs 02.50.Le, 87.23.Ge, 07.05.Tp
\end{abstract}

\section{Introduction}

Games and economic models are more interrelated than one can
imagine~\cite{
	Kreps1990Book}.
This is also the case for social interactions.
A simplistic virtual setting for simulating a trust
in an e-commerce
setting,
would be the Iterated Prisoners Dilemma game which is, by its nature,
very related to the evolution of
a trust~\cite{
	Axelrod2006Book,
	Tesfatsion1997}.
Each transaction in an e-commerce setting can be viewed as a round in an iterated prisoner's dilemma game.
Adherence to electronic contracts or providing services with good quality can be considered as
cooperation while the temptation to act deceptively for immediate gain can be considered as deception.

Economy is the study of how to allocate scarce resources.
According to Davenport,
the scarcest resource of today is nothing but attention~\cite{Davenport2001Book}.
Attention scarcity is first stated by Herbert Simon.
He says that, ``What information consumes is rather obvious:
it consumes the attention of its
recipients''~\cite{
	Simon1971}.
The new digital age
has come
with its vast amount of immediately available information
that
exceeds our information processing power.
Thus,
attention scarcity
is a natural consequence of
huge amount of information.
Attention is very critical to any kind of interaction,
especially in the era of digital technologies.
Conventional Economy has been transforming itself to the
Attention Economy~\cite{
	Goldhaber1997attention,
	Davenport2001Book,
	Falkinger2007,
	Falkinger2008}.
Games should do the same.
Little work has been done on games with limited attention.
How does attention scarcity effect a game?
We will discuss attention games in a specific context of Iterated Prisoners Dilemma.

\subsection{Iterated Prisoners Dilemma game}

Prisoners Dilemma game is one of the commonly studied social
experiments~\cite{
	Axelrod2006Book,
	Axelrod1997Book,
	Axelrod1981Science,
	Rapoport1965Book,
	Kollock1998,
	Tesfatsion1997}.
Two players should simultaneously select one of the two actions: cooperation or defection,
and play accordingly with each other.
Dependent on their choices,
they receive different payoffs as seen in
\reffig{fig:payoff}.

Payoff matrix can be described by the following simple rules.
In the case of mutual cooperation,
both players receive the
\emph{reward} payoff, $R$.
If one cooperates, while the other defects,
cooperator gets  the
\emph{sucker's} payoff, $S$
while the defector gets
\emph{temptation} payoff, $T$.
In the case of mutual defection, both get the
\emph{punishment} payoff $P$.
Payoff matrix should satisfy
the inequality
$S < P < R < T$ and
the additional constraint
$T + S < 2 R$ for repeated interactions.
Rationality leads to defection, because
$R < T$ and
$S < P$ makes defection better than cooperation.
But, at the same time,
$P < R$ implies that
mutual cooperation is superior to
mutual defection.
So, rationality fails and
this situation is referred to as a dilemma.

\begin{figure}[!t]
\begin{center}
	\begin{tabular}{c|c|c}
		&Cooperate
		&Defect \\
		\hline
		Cooperate
		&$R$, \; $R$
		&$S$, \; $T$ \\
		\hline
		Defect
		&$T$, \; $S$
		&$P$, \; $P$ \\
	\end{tabular}
\caption{
	Payoff matrix.
	We use 
	$T = 5$,
	$R = 3$,
	$P = 1$, and
	$S = 0$.
}
\label{fig:payoff}
\end{center}
\end{figure}

It is well known that
the defection is the individually reasonable behavior
that leads to a situation
in which everyone is worse
off~\cite{
	Axelrod2006Book}.
On the other hand, cooperation results in the maximization of the joint
outcomes~\cite{
	Kollock1998}.

If two players play prisoners dilemma more than once and
they remember previous actions of their opponent and change their strategy accordingly,
the game is called
\emph{Iterated Prisoners Dilemma} (\emph{IPD})~\cite{
	Rapoport1965Book}.
Despite its level of abstraction,
a large variety of situations starting from daily life
(i.e., stop or go on when the red light is on?)
to socio-economic relations
(i.e., fulfill or renege on trade obligations?)	
may be represented as an IPD game.
It is shown that repeated encounters
between the same individuals
foster cooperation.
This is often referred to as the \emph{shadow of the future}.
If individuals are likely to interact again in the future,
this allows for the return of an altruistic act~\cite{
	Axelrod2006Book,
	Axelrod1997Book}.

\subsection{Attention in games}

In general, a player is not capable of knowing all the players in an interacting environment and usually acts based on a limited information.
One reason could be
the huge number of players,
or another could be that
the players may have a very limited memory size to be informed of all the
others~\cite{
	Bingol2005LNCS,
	Bingol2008PRE}.
For example,
in real life,
a market has a few market leaders and many small brands
whose number, in general, is simply too large for consumer to remember all of them.
Therefore, a consumer can only have access to a limited number of service providers.
The essence of any game is
to interact with other players
and get a chance
to improve the payoff one gets.
To interact with others,
one should first capture their attention in a positive manner.
When we give our attention to something,
we always take it away from something else.
We can think of having attention as owning a kind of property.
This property is located in the memory of a player.

\subsection{IPD game under limited attention}

In many studies related to IPD game,
it is assumed that
there exists enough memory to remember
all the previously encountered players and their actions.
Memory is an important aspect,
because knowing the identity and history of an opponent
allows one to respond in an appropriate manner.
We use the term \emph{limited attention}
to indicate the existence of an upper bound on
how many distinct encounters are remembered by a player.
We ask the following reasonable question,
as in  \refcite{
	Bingol2008PRE},
what if the memory size is limited?
The same question can be reformulated as follows:
what if attention capacity is limited?
In this study, we introduce attention capacity as an important parameter to investigate the dynamics of the mentioned game.

\section{The model}

Researcher Tesfatsion introduced the notions of choice and refusal into IPD
games~\cite{
	Tesfatsion1997}.
In order to choose or refuse an opponent,
players should be able to remember
the identity of each player and their past behaviors.
It is known that the
choice helps players to find cooperation
while refusal lets them escape from
defection~\cite{
	Tesfatsion1997}.
In our very simplistic model,
we consider that there exist two type of players:
\emph{cooperators}, who always cooperate, and
\emph{defectors}, who always defect.
We combine these pure strategies with
a simple choice-and-refusal rule:
If a player knows that the opponent is a defector,
then he or she refuses to play.
Otherwise he or she plays.

Each round of the IPD game consumes a limited attention of its players.
We assume that every player has the same \emph{attention capacity} $M$.
When a player encounters an opponent,
he stores the necessary information related to the opponent's
action in his memory.
After playing with $M$ different opponent,
the attention capacity fills up.
As the player encounters more opponents,
he will have the problem of attention scarcity.
He has to forget the previously encountered ones.
To use ones memory efficiently,
one needs to decide whom to forget?
In this respect, in section~\ref{sec4.1} we will discuss
5 different attention allocation strategies.
Like the rest of the literature,
we focus on the conditions under which
``cooperative move'' becomes more favorable.
However, our research considers
that the game takes place in a world with a limited attention.

The personality of a player (cooperator or defector)
is randomly set.
Remember that once the personality is set, it never changes.
In each iteration, two individuals are randomly chosen to play the game.
In this respect,
there is no spatial pattern.
One considers that the underlying interaction graph is a complete graph.

Let $C$ and $D$ denote the sets of cooperator and defector players, respectively.
Let $\mathcal{N}$ denote the set of all players,
that is,
$\mathcal{N} = C \cup D$.
The number of defectors
is denoted by $|D|$.
Thus, the remaining $|C| = N-|D|$ players are the cooperators,
where $N = |\mathcal{N}|$.
We define our model parameters
\emph{attention capacity ratio}
and
\emph{defector ratio}
as
$\mu = {M}/{N}$
and
$\delta= {|D|}/{N}$, respectively.
Hence,
we have
$0 \leqslant \mu \leqslant 1$ and
$0 \leqslant \delta \leqslant 1$.

We use the de facto payoff values of
$T = 5$,
$R = 3$,
$P = 1$, and
$S = 0$
throughout this study.

\section{Evaluation metrics}

Social welfare
can be measured by
the average payoff of players.
The payoffs of all the encounters are added up to have the final outcome of each player.
To make a comparison between the defectors and the cooperators, we take the average outcome of each.
Let $c_{i}$ and $d_{i}$ be the numbers of games,
where the player $i$ plays with cooperators and defectors, respectively.
We use the payoff matrix given in \reffig{fig:payoff}
to calculate the total payoff of the player $i$ as follows:
\[
	\payoff(i)
	=
	\begin{cases}
		R c_{i} + S d_{i},
		&i \in C, \\
		T c_{i} + P d_{i},
		& \text{otherwise}.
	\end{cases}
\]
We evaluated our results by a comparison between
the average performances of the cooperators and
the average performances of the defectors.
Our performance metrics are as follows:
\[
	\bar{P}_{C}= \frac{1}{|C|} \sum_{i \in C} \payoff(i)
	\text{ \; and \; }
	\bar{P}_{D}= \frac{1}{|D|} \sum_{i \in D} \payoff(i).
\]
Although further investigations call for simulations,
some analytical investigation of average performances is possible.

\subsection{Cooperator's average performance}
\label{sec:PCAnalytic}

Cooperator's average performance of $\bar{P}_{C} $ can be analytically found.
For a cooperator, to play with a defector means no gain,
since sucker's payoff is equal to zero,
that is,
$S=0$.
$\bar{P}_{C} $ can only increase if two cooperators play a round with each other.
When two cooperators are selected to play with each other,
each cooperator gets $R=3$ points.
The probability of matching two cooperators is equal to $(1-\delta)^2$.
Among $\mathcal{T} = \tau {N^2}/{2}$ rounds,
only $(1 - \delta)^2 \mathcal{T}$ of them is expected to pass between two cooperators.
As a result, $|C| = (1-\delta)N$ cooperators share
$ (R+R) (1-\delta)^2\tau {N^2}/{2}$ payoffs.
In other words,
\[
	\bar{P}_{C}
	= \frac
		{2 R (1 - \delta)^2 \tau \frac{N^2}{2}}
		{(1 - \delta) N}
	= R (1-\delta) \tau N.
\]
Without any further investigation,
we can conclude that
increasing $\tau $,  $N$ and $R$ is favorable for
$\bar{P}_{C}$ while increasing $\delta$ is not.
Note that
neither attention capacity $M$
nor any attention allocation strategy
has effect in this setting.
If the population is composed of only cooperators,
that is $|C| = N$ and $\delta = 0$, $\bar{P}_{C}$ will be $R \tau N$.

\subsection{Defector's average performance}

Due to the choice and refusal rule,
if an opponent is known to be a defector,
no player plays with him.
Therefore,
in order to obtain
the defector's average performance of $\bar{P}_{D}$,
we need
the probability of
a defector $j \in D$
to be unknown by
player $i \in \mathcal{N}$.
This probability cannot be analytically found
except for the special cases of
players without memory
and
players with unlimited memory.

\subsubsection{Players without memory}
\label{sec:PDwMu0}

When players have no memory,
i.e., attention capacity is zero,
they are totally forgetful and remember nothing.
Note that this case actually corresponds to a player playing prisoners dilemma
without realizing that they are playing repeatedly.
As a result, players continue to play with defectors
in spite of the choice and refusal rule.
The probability of matching a defector with a cooperator
is equal to
$2 \delta (1 - \delta)$
while matching the two defectors is equal to
$\delta^2$.
Therefore, for a special case of $\mu = 0$,
we have
\[
	\bar{P}_{D}
	= \frac
		{\left[ T \; 2 \delta (1 - \delta) + 2P \; \delta^2 \right] \tau \frac{N^2}{2}}
		{\delta N}
	=\left[ T \;(1-\delta) + P \; \delta \right]\tau N
	= (5 -4\delta)\tau N
\]
for $T = 5$ and $P = 1$.
We observe that increasing the number of defectors is not
favorable even for defectors.
Nevertheless,
it is easy to verify that
for $\mu = 0$, $\bar{P}_{D} $ is always greater than $\bar{P}_{C}$
which can be stated as
\emph{defection is a favorable action against the players with no memory}.

\subsubsection{Players with unlimited memory}
\label{sec:PDwMu1}

For a special case of $M \geqslant N$,
the players are no longer forgetful
and they are able to remember each opponent's last action.
Due to the choice and refusal system,
any defector can play at most $|C|$ rounds with cooperators and
$|D| -1$ rounds with defectors.
Therefore, for a sufficiently large $\tau$,
we have
\[
	\bar{P}_{D}
	= T  |C| + P  (|D| - 1)
	= (P - T) |D| + T N - P\,.
\]
We can conclude that
as we increase the number of defectors in this setting,
the average payoff of the defectors again decreases.

\section{Simulations}

The dynamics of a system is further investigated by simulation while
the attention capacity ratio $\mu$ and
the defectors ratio $\delta$ vary.
The model is simulated for every possible attention capacity values of $M$ (from 0 to $N$)
and for every possible number of defectors (from 0 to $N$).
We study a population of $N = 100$.

The number of iterations, $\mathcal{T}$,
is another critical issue.
It is set to
$\mathcal{T} = \tau \times {N^{2}}/{2}$
since there are ${N \choose 2}$ pairs,
where
$\tau$, being the third model parameter, is the number of plays for a pair of players.
Note that, when $\tau=1$,
no two players are expected to meet again during the simulation.
This situation corresponds to a non-iterated version of the game.
In order to see the effect of time, $\tau$ is set to 2 and 5.
The results were averaged over 20 independent realizations
for every combination of parameter values.

\subsection{Attention allocation strategies\label{sec4.1}}

Some people are positive and remember only good memories.
On the contrary, some remember bad events and
live to get their revenge.
Motivated by these,
we make a comparison of 5 simple attention allocation strategies
based on forget mechanisms:
(i)~Players that prefer to
forget only cooperators,
denoted by {FOC}.
(ii)~Players that prefer to
forget only defectors,
denoted by {FOD}.
(iii)~When players have no preference,
they can select someone,
uniformly at random, to forget.
We call this strategy as {FAR}.
(iv)~Players may also prefer to use coin flips to decide which type,
namely, cooperators or defectors,
of a player to forget.
Once the type is decided, someone among this type is randomly selected and forgotten.
Let {FEQ} denote this ``equal probability'' to types approach.
(iv)~
If the knowledge of which type has the majority is available,
this extra information can be used in devising a strategy.
One possible effective strategy could be
to assume that the opponent is of the type of majority,
hence, pay attention to the minorities only.
That is, one prefers to forget the majority which we call {FMJ} strategy.

We investigate the average performances of cooperators and defectors
when they use the same strategy.

\section{Observations}
\label{sec:Observations}

\begin{figure*}
\begin{center}
\includegraphics[width=1.0\textwidth]{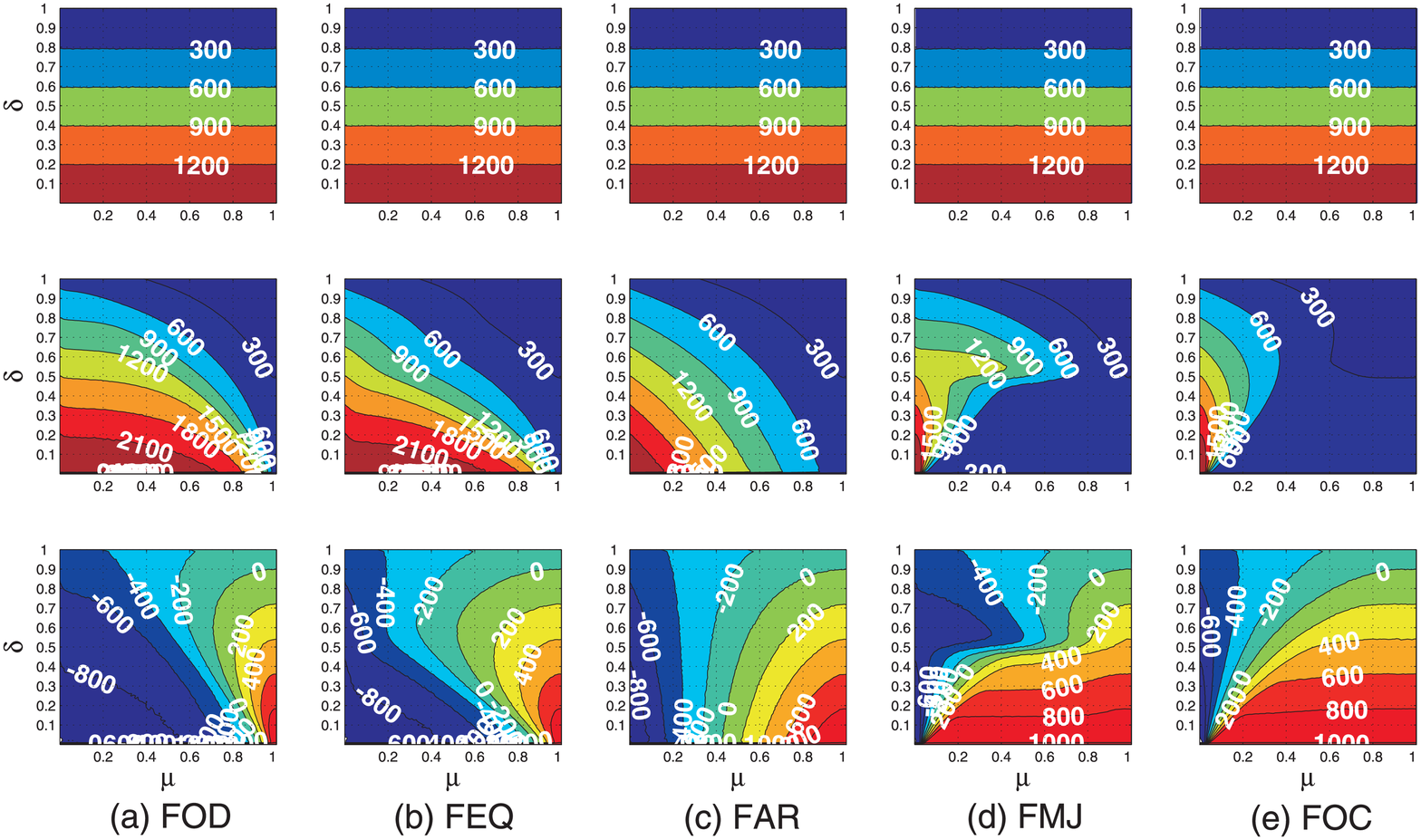}
	\caption{
		(Color online)
		Average performances as a function of
		attention capacity ratio $\mu$ and
		defector ratio $\delta$.
		The columns represent five strategies.
		The rows represent
		$\bar{P}_{C}$,
		$\bar{P}_{D}$ and
		$\bar{P}_{C} - \bar{P}_{D}$
		values, respectively.
	}
	\label{fig:AllStrategies}
\end{center}
\end{figure*}

In this section, for a more general view, we present our observations
based on our simulation data.
With our essential parameters of
$\mu$,
$\delta$,
 and
$\tau$
along with the different attention allocation strategies,
we can determine the conditions under which cooperation is more favorable
than defection.

Simulation results for various values of
attention capacity ratio $\mu$ and
defector ratio $\delta$  are given in
\reffig{fig:AllStrategies}.
Columns of
\reffig{fig:AllStrategies} correspond to five strategies.
Within a column,
the top plot provides the average performance of cooperators,
$\bar{P}_{C}$,
as a function of $\mu$ and $\delta$.
Similarly, the middle plot gives the average performance of defectors.
The bottom plot is the difference of the averages.
Note that being a cooperator is better
when $\bar{P}_{C} - \bar{P}_{D} > 0$.
For the sake of comparison,
$\bar{P}_{C}-\bar{P}_{D} = 0$ curves
for different attention allocation strategies are superposed in
\reffig{fig:superpositionTau5}.

\subsection{Average performance of cooperators}

Findings from the first row of
\reffig{fig:AllStrategies} are as follows:
(i)~Interestingly, cooperator's average payoff does not significantly change
neither by attention capacity ratio nor by attention allocation strategy.
(ii)~However, the defector ratio has a negative effect on the average performances of cooperators.
Our analytical explanation given in \refsec{sec:PCAnalytic}
is in agreement with these findings.
For any $\delta$ values,
$\bar{P}_{C} = R (1 - \delta) \tau N$
gives exactly the same results seen
in the first row of
\reffig{fig:AllStrategies}.

\subsection{Average performance of defectors}

The second row of
\reffig{fig:AllStrategies} can be interpreted as follows:
(i)~Greater attention capacity,
i.e., an increase in $\mu$,
helps players to remember the defectors.
As a result,
defectors experience social isolation and
their average payoff severely diminishes.
(ii)~An increase in the number of defectors,
i.e., an increase in $\delta$,
leads a competition among them.
Thus, defectors' average payoff again diminishes.
(iii)~Note that all five plots are in agreement with
our discussion in \refsec{sec:PDwMu0} and \refsec{sec:PDwMu1}
for special cases of $\mu=0$ and $\mu=1$.

\subsection{Attention boundaries}

\begin{figure*}
\begin{center}
	\subfigure[$\tau=2$]{
\includegraphics[width=0.48\columnwidth]{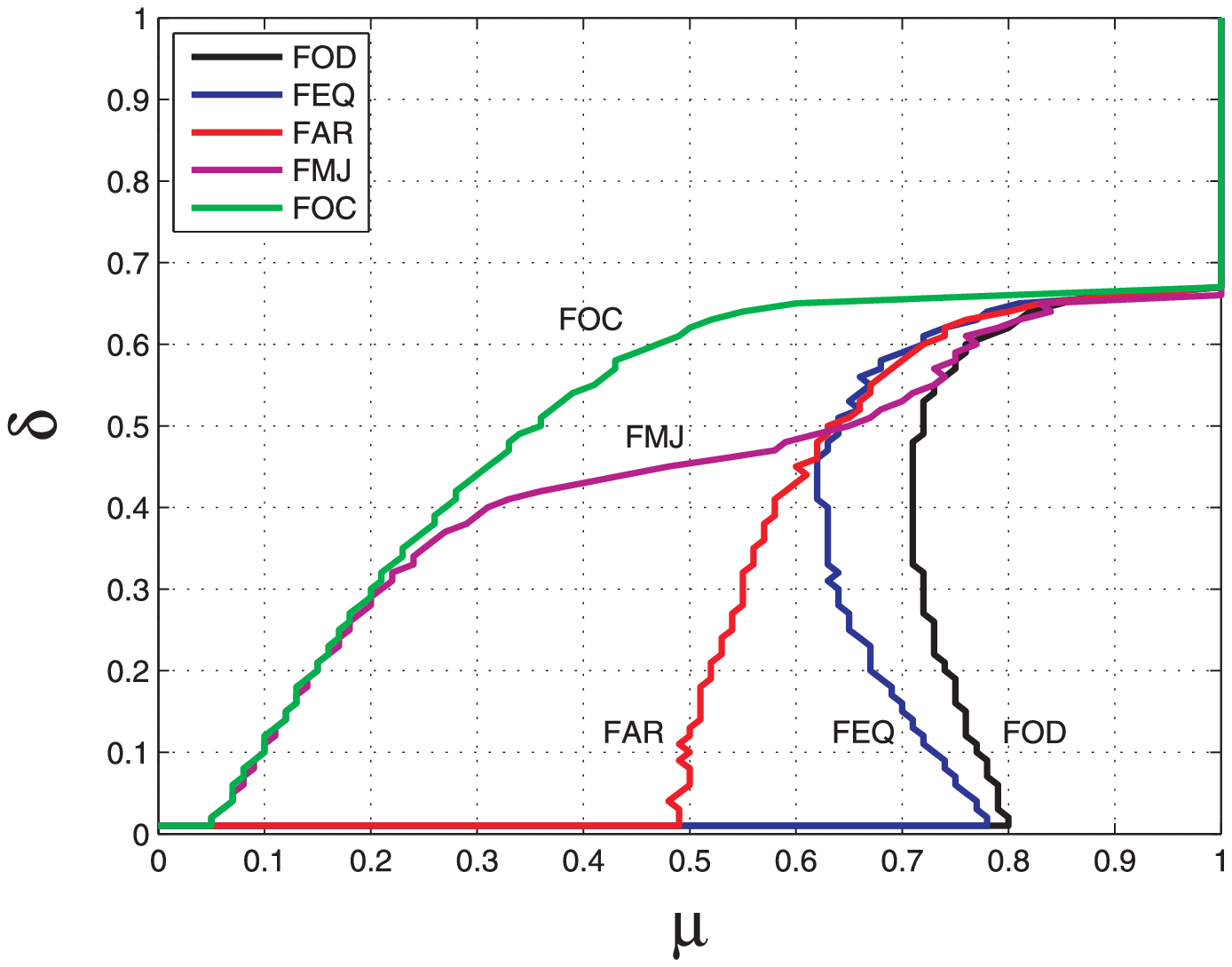}
		\label{fig:superpositionTau2}
	}
	\subfigure[$\tau=5$]{
\includegraphics[width=0.48\columnwidth]{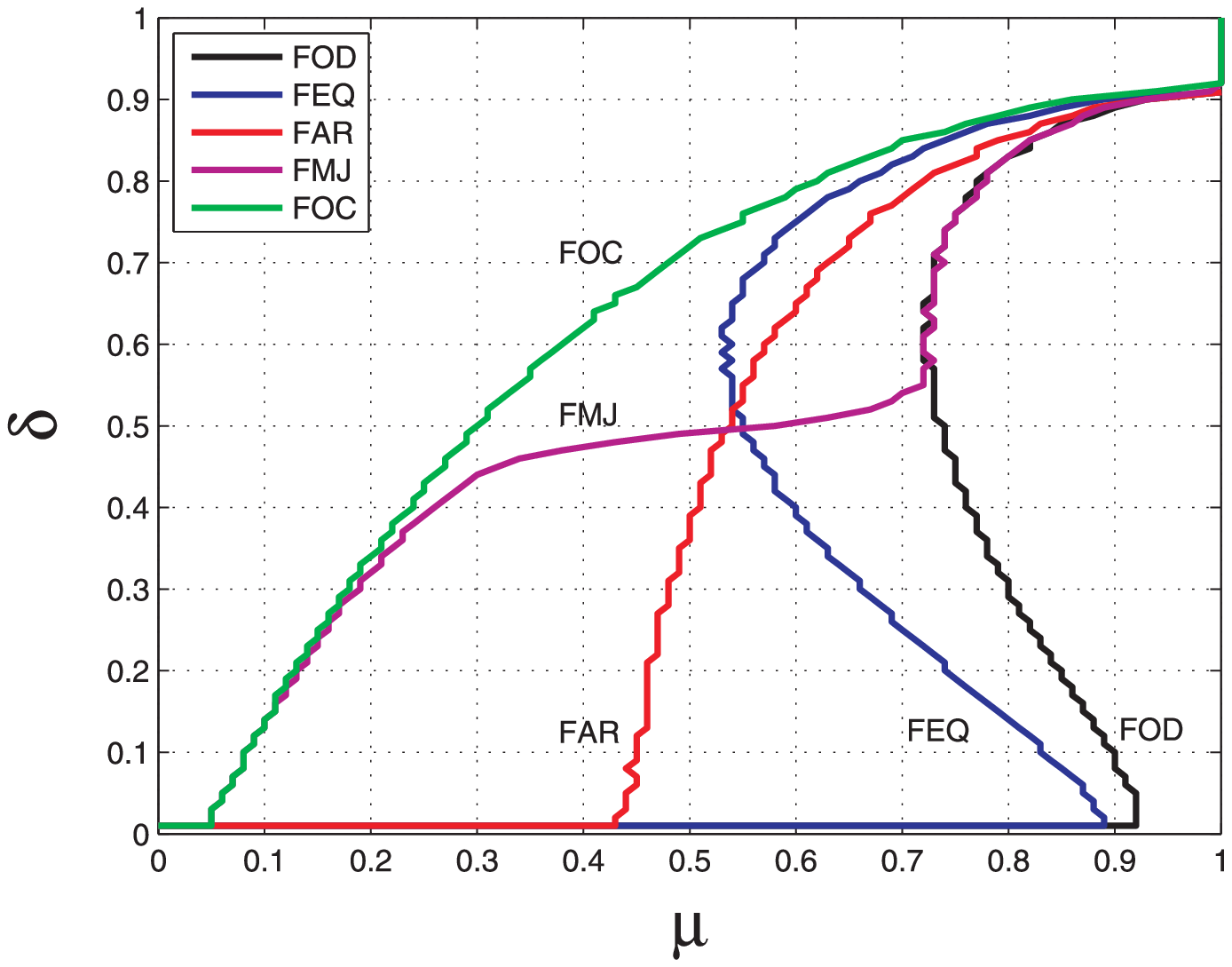}
		\label{fig:superpositionTau5}
	}
	\caption{
		(Color online)
		Attention boundaries of
		different allocation strategies
		are visualized in the same figure for the sake of comparison.
	}
	\label{fig:superposition}
\end{center}
\end{figure*}

We refer to the $\bar{P}_C -\bar{P}_D = 0$ contour lines,
seen in the third row of
\reffig{fig:AllStrategies},
as the \emph{attention boundaries}.
An attention boundary
determines a favorable action.
If a pair of $(\mu, \delta)$ remains inside the attention boundary,
it means $\bar{P}_C - \bar{P}_D > 0$ and cooperation is a favorable action,
otherwise defection is a favorable action.
Attention boundaries
for five different attention allocation strategies seen in
\reffig{fig:AllStrategies},
are visually superposed in
\reffig{fig:superpositionTau5}
for the sake of comparison.

For a given defector ratio,
we observe that there is a critical threshold for attention capacity,
below which defection is advantageous,
and above which cooperation becomes a favorable action.	
With lesser attention capacity, defectors can be easily overlooked.
Greater attention capacity along with the
choice-and-refusal rule do not let defectors improve their payoffs.
Due to the degrading of defector's performance,
the average payoff of cooperators manages to
exceed that of defectors
when players have a greater attention capacity.

\subsection{Attention allocation strategies}

We consider a strategy \emph{better}
if it has a larger area,
where cooperators are doing better than defectors,
in the $\mu, \delta$ plain.
That is, a better strategy has more $(\mu, \delta)$ pairs below its attention boundary.
From this perspective, the best strategy is FOC, and the worst one is FOD.
All the remaining strategies are located in between these two strategies.

The forget majority, FMJ, is a mixed strategy.
When $0.5 < \delta$,
defectors are the majority and
FMJ acts as if they forget only the defectors.
When $\delta < 0.5$,
cooperators are the majority.
Thus FMJ switches to forget only cooperator.
Therefore, its plot is similar to that of FOD for $0 < \delta < 0.5$
and that of FOC
for $0.5 < \delta < 1$.
FMJ strategy can be put differently as allocation of the minority.
One can think that this strategy is better than the rest,
since scarcity, in general, triggers the perception of greater importance.
Nevertheless, \reffig{fig:superpositionTau5} is against this intuition.
The optimal strategy is to forget only the cooperators.
By doing so, players manage to allocate their memories only for defectors.
In other words, they keep their enemies closer.
Thus, they become more prudent to the defectors.
On the other hand,
forgetting defectors seems to be the
most wasteful and carefree attention consuming habit.
We observe that
the necessary information for refusing the defectors
is dismissed
while applying the FOD strategy.

The critical value of $\delta = 0.5$ determines which strategy is superior,
except for the two extreme strategies of FOD and FOC.
FEQ does better than FAR when
$0.5 < \delta$ and FAR does better than FEQ when
$\delta < 0.5$.
Even if FAR strategy seems identical to FEQ strategy,
there exists a slight difference between them.
Notice that, forgetting at random depends on the content of the memory,
while forgetting with equal probability does not.
Higher defector's ratio, that is $0.5 < \delta$, causes
one to encounter more defectors.
In that case, memories of the players would be plentiful with defective experiences.
Thus, forgetting at random would be more biased towards FOD.
Similarly,
forgetting at random would be more biased towards FOC
when $\delta < 0.5$.

\subsection{Effect of time}

Literature on IPD game suggests that
as the number of iterations increases, the
cooperative behavior also increases among the players~\cite{
	Axelrod2006Book,
	Axelrod1997Book}.
This is also verified by our simulations.
		The shadow of the future
		can be quantified by the parameter of  $\tau$.
A short shadow of the future (lesser $\tau$), hinders the detection of the defectors.			
When the future of the shadow is longer,
lesser attention capacity would be sufficient for cooperators to beat the defectors.
As $\tau$ increases,
defector's performance gets worse
in comparison with cooperators.
Attention boundaries obtained by setting $\tau=2$ and $\tau=5$
are given in \reffig{fig:superpositionTau2} and \reffig{fig:superpositionTau5}, respectively.
The area inside the attention boundaries
is much larger in
\reffig{fig:superpositionTau5}
than
in \reffig{fig:superpositionTau2}.
This finding suggests that
the shadow of the future fosters cooperation.

\section{Conclusions}

We observe
that as the proportion of the defectors increases,
the average payoff for any player decreases.
On the other hand,
an increase in the attention capacity
has different outcomes for cooperators and defectors.
As attention capacity increases,
the change in the
cooperators overall performance is almost negligible,
but the defectors' performance significantly diminishes.
The rule of choice-and-refusal plays an important role in this situation.
Nevertheless,
it is worth pointing out that
even the choice-and-refusal alone
cannot fulfill the desired goal
without passing some threshold value of
attention capacity.
As the attention capacity increases,
or the shadow of the future gets longer,
the detection of the defectors gets feasible,
consequently the defectors face a social isolation due to the rule of choice-and-refusal.
As a result, the cooperators' performance
exceeds the defectors' performance.
Thus, cooperation becomes a favorable action.
This work demonstrates that in order to beat a defection,
players do not need a full memorization of each action of all opponents.
This finding is really important especially in the world of a limited attention.
We also investigate five different attention allocation strategies and
we find out that the best strategy is ``forgetting only the cooperators''.
By applying this strategy,
one becomes more prudent to the deceptive actions.
In conclusion, attention should be selective,
and it should be directed towards the defectors
and towards their defective moves.

In the present work,
players are pure cooperators or pure defectors.
They never change their character.
Various forgetting strategies are investigated but
both cooperators and defectors use the same strategy in a game.
The situation of cooperators using one strategy and defectors another is left for the  future work.
It would be also interesting to study the effect of a biased payoff matrix.
As a future work,
we plan to investigate other means for fostering cooperation,
even in the conditions of attention scarcity.
To achieve this goal,
we can make use of other experiences by taking recommendations to determine with whom to play.
But from whom to take advice is very critical
and must be well studied to clarify which collaboration strategy is better.
We will also extend our work to the mixed strategies for interaction,
such as ``mostly defect'' and ``mostly cooperate''.

\clearpage



\ukrainianpart

 \title{Ітерована дилема  в'язня з обмеженою увагою}
 \author{
  У. Сетін\refaddr{label1,label2},
  Г.O. Бінгол\refaddr{label1}}
 \addresses{
  \addr{label1} Факультет комп'ютерної інженерії, Університет Богазічі, Стамбул, Туреччина
  \addr{label2} Факультет комп'ютерної інженерії, Стамбульський університет Гелісім, Стамбул, Туреччина
 }

 \makeukrtitle

 \begin{abstract}
 \tolerance=3000%
 Як дефіцит уваги впливає на результати гри?
 Ми представляємо наші результати на прикладі гри Ітерована дилема в'язня (Iterated Prisoners Dilemma (IPD)), в якій
 гравці можуть погоджуватися чи відмовлятися грати зі своїм партнером.
 Ми вивчаємо вплив розміру пам'яті на визначення відповідного партнера для взаємодії.
 Ми досліджуємо умови, при яких ймовірніше стати партнерами ніж перебіжчиками.
 Ця робота демонструє, що для перемоги над дезертирством гравці не потребують повного запам'ятовування кожної дії всіх опонентів.
  Для того, щоб перемогти перебіжчиків існує критичний поріг здатності уваги.
 Цей поріг залежить не тільки від частки перебіжчиків в популяції, але також від стратегії розподілу уваги гравців.
 \keywords
  дефіцит уваги, взаємодія, вплив розміру пам'яті, ітерована дилема в'язня, соціальні та економічні моделі
 \end{abstract}

 \end{document}